\documentclass[11pt]{article}
\usepackage{graphicx}
\usepackage{relsize}
\usepackage{rotating}

\newcommand{\BABARPubYear}    {04}
\newcommand{\BABARConfNumber} {51}
\newcommand{\SLACPubNumber} {10709}

\input babarsym

\input mymacros.tex

\newcommand{\lumi}    {221.4\invfb}
\newcommand{\lumion}  {205.3\invfb}
\newcommand{\lumioff} {16.1\invfb}
\newcommand{\milliontaudecays} {390}
\newcommand{\taulhhlimits} {\ensuremath{(0.7-4.8)\times 10^{-7}}}

\long\def\inst#1{\par\nobreak\kern 4pt\nobreak
    {\it #1}\par\vskip 10pt plus 3pt minus 3pt}

\begin{document}
{\pagestyle{empty}

\begin{flushright}
\babar-CONF-\BABARPubYear/\BABARConfNumber \\
SLAC-PUB-\SLACPubNumber \\
\end{flushright}

\par\vskip 5cm

\begin{center}
\Large \bf \boldmath Search for Lepton-Flavor Violation in the Decay \taulhh
\end{center}
\bigskip

\begin{center}
\large The \babar\ Collaboration\\
\mbox{ }\\
\today
\end{center}
\bigskip \bigskip

\begin{center}
\large \bf Abstract
\end{center}
A search for the lepton-flavor-violating decay of the 
tau into one charged lepton and two charged hadrons 
has been performed using \lumi\ of data collected at an \epem\
center-of-mass energy around 10.58\gev with the \babar\ detector 
at the \pep2\ storage ring.
In all 14 decay modes considered, the numbers of events found 
in data are compatible with the background expectations.
Upper limits on the branching fractions are set in the range 
\taulhhlimits\ at 90\% confidence level.
All results are preliminary.

\vfill
\begin{center}

Submitted to the 8$^{\rm th}$ International Workshop on Tau Lepton Physics, 
Tau 04,\\
14 September---17 September 2004, Nara, Japan

\end{center}

\vspace{1.0cm}
\begin{center}
{\em Stanford Linear Accelerator Center, Stanford University, 
Stanford, CA 94309} \\ \vspace{0.1cm}\hrule\vspace{0.1cm}
Work supported in part by Department of Energy contract DE-AC03-76SF00515.
\end{center}

\newpage
} 

%
%
\renewcommand{\thefootnote}{\fnsymbol{footnote}}

\begin{center}
\small

The \babar\ Collaboration,
\bigskip

%
B.~Aubert,
R.~Barate,
D.~Boutigny,
F.~Couderc,
J.-M.~Gaillard,
Y.~Karyotakis,
J.~P.~Lees,
V.~Tisserand,
A.~Zghiche
\inst{Laboratoire de Physique des Particules, F-74941 Annecy-le-Vieux, France }
A.~Palano,
A.~Pompili
\inst{Universit\`a di Bari, Dipartimento di Fisica and INFN, I-70126 Bari, Italy }
% ...

\end{center}\newpage

\renewcommand{\thefootnote}{\arabic{footnote}}
\setcounter{footnote}{0}

\section{INTRODUCTION}
\label{sec:Introduction}

Lepton-flavor violation (LFV) involving charged leptons has 
never been observed, and stringent experimental limits 
exist from muon branching fractions:
$\BR(\mmu\to\electron\gamma) < 1.2 \times 10^{-11}$ \cite{brooks99}
and $\BR(\mmu\to\electron\electron\electron) < 1.0 
\times 10^{-12}$ \cite{sindrum88} at 90\% confidence level (CL).
Recent results from neutrino oscillation experiments \cite{neut} 
show that LFV does indeed occur, although the branching fractions 
expected in charged lepton decays due to neutrino mixing alone 
are probably no more than $10^{-14}$ \cite{pham98}.

In tau decays, the most stringent limit on LFV is 
$\BR(\taum\to\eemw) < 1.1 \times 10^{-7}$ 
at 90\% CL \cite{babar03}.
Many extensions to the Standard Model (SM), particularly models 
seeking to describe neutrino mixing, predict enhanced LFV in tau
decays over muon decays with branching fractions from 
$10^{-10}$ up to the current experimental limits \cite{ma02}.
Observation of LFV in tau decays would be a 
clear signature of non-SM physics, while improved 
limits will provide further constraints on theoretical models.

This paper presents preliminary results of a search for 
the decays\footnote{
Throughout this paper, charge conjugate decay modes also are implied.}
$\taulhh$ where $\ell$ represents an electron or muon 
and $h$ represents a pion or kaon.
In total there are 14 distinct final states consistent with
charge conservation.
The best existing limits on the branching fractions for these
decay modes currently come from CLEO, and range from $(2-8)\times10^{-6}$
at 90\% CL \cite{cleolhh}.

\section{THE \babar\ DETECTOR AND DATASET}
\label{sec:babar}
The data used in this analysis were collected with the \babar\ detector
at the \pep2\ asymmetric-energy  $e^+e^-$ storage ring.
The data sample consists of \lumion\ recorded at
$\sqrt{s} = 10.58 \gev$ and \lumioff\ recorded at
$\sqrt{s} = 10.54 \gev$.
With an estimated luminosity-weighted cross section for tau pairs
of $\sigma_{\tau\tau} = (0.89\pm0.02)$ nb \cite{kk},
this data sample contains over \milliontaudecays\ million tau decays.

The \babar\ detector is described in detail elsewhere~\cite{detector}.
Charged-particle (track) momenta are measured with a 5-layer
double-sided silicon vertex tracker and a 40-layer drift chamber 
inside a 1.5-T superconducting solenoidal magnet.
An electromagnetic calorimeter (EMC) consisting of 6580 CsI(Tl) 
crystals is used to identify electrons and photons,
a ring-imaging Cherenkov detector and energy loss in the drift chambers are used to identify
charged hadrons, 
and the instrumented magnetic flux return (IFR) is used to
identify muons.

A Monte Carlo (MC) simulation of LFV tau decays
is used to study the performance of this analysis.
Simulated tau-pair events including higher-order radiative
corrections are generated using \kktwof \cite{kk}
with one tau decaying to one lepton and two hadrons with a 
3-body phase space distribution, while the other tau decays 
according to measured rates \cite{PDG} simulated with 
\tauola \cite{tauola}.
Final state radiative effects are simulated for all decays 
using \photos \cite{photos}.
The detector response is simulated with \mbox{\tt GEANT4}~\cite{geant},
and the simulated events are then reconstructed in the same 
manner as data.

\section{ANALYSIS METHOD}
\label{sec:Analysis}
The analysis procedure is similar to that used in our
published \taullls\ analysis \cite{babar03}. 
All possible lepton and hadron combinations consistent with charge
conservation are considered, leading to fourteen distinct
decay modes (\EKKr, \EKPr, \EPKr, \EPPr, \MKKr, \MKPr, \MPKr, \MPPr, 
\EKKw, \EKPw, \EPPw, \MKKw, \MKPw, \MPPw).
The signature of this process is three charged 
particles, one identified as either an electron or muon,
and each of the other two identified as either a pion or kaon,
with an invariant mass and energy equal to that of the parent 
tau lepton.

\subsection{SELECTION}
Candidate signal events in this analysis are required
to have a ``1-3 topology,'' where one tau decay yields three
charged particles (3-prong), while the second tau
decay yields one charged particle (1-prong).
Four well reconstructed tracks are required 
with zero net charge, 
pointing towards a common region consistent with 
\tautau{} production and decay.
The event is divided into two hemispheres using the
plane perpendicular to the thrust axis, calculated from the 
observed tracks and unassociated EMC energy deposits 
in the center-of-mass (CM) frame. 
One hemisphere must contain exactly one track while the
other must contain exactly three, defining the 1-3 topology.
Pairs of oppositely charged tracks identified as
photon conversions in the detector material with 
an \epem invariant mass below 30\mevcc are ignored.

One of the charged particles found in the 3-prong 
hemisphere must be identified as either an electron
or muon candidate.
Electrons are identified using the ratio of
EMC energy to track momentum $(E/p)$, the ionization 
loss in the tracking system $(\dedx)$, and the shape of the shower
in the EMC.
Muons are identified by hits in the IFR
and small energy deposits in the
EMC.
Muons with momentum less than $0.5\gevc$ cannot be identified
because they do not penetrate far enough into the IFR.
Each of the other two charged particles found in the 3-prong
hemisphere must be identified as either a pion or a kaon.

The particle identification requirements alone
are not sufficient to suppress certain 
backgrounds, particularly those from light quark \qqbar\ 
production and SM \tautau\ events, therefore additional selection 
criteria are required.
The selection requirements, most of which are the same for
all 14 decay modes, are as follows:
\begin{itemize}
\item no neutral clusters (photon candidates) with energy
      $(E_{\gamma} > 100 \mev)$ in the EMC. 
      This restriction removes mostly \qqbar\ backgrounds
      and some SM tau-pair events;
\item the polar angle of the missing momentum in the lab frame,
      $\Theta_{\mathrm{miss}}$, is required to be in the range 
      $0.25\ \mathrm{rad} < \Theta_{\mathrm{miss}} < 2.4\ \mathrm{rad}$.
      This cut is effective at reducing two-photon
      and Bhabha backgrounds;
\item the total transverse momentum in the event in the CM frame,
      $p_T^{\mathrm{CM}}$, must be greater than 0.2 \gevc. 
      The cut on $p_T^{\mathrm{CM}}$ is also
      effective against Bhabha and two-photon events;
\item the mass of the one-prong hemisphere, $m_{1pr}$, calculated from the 
      four-momentum of the track in the 1-prong hemisphere and the missing 
      momentum in the event, is required to satisfy
      $0.6\gevcc\ < m_{1pr} < 1.9\gevcc$ for \ehh\ final
      states and $0.8\gevcc\ < m_{1pr} < 1.9\gevcc$ for \mhh\ final states.
      The one-prong mass requirement is particularly effective at removing 
      \qqbar\ backgrounds as well as the remaining two-photon contribution;
\item to remove any remaining Bhabha background, the 
	  momentum of the one-prong track in the CM frame, 
	  $p_{1pr}^{\mathrm{CM}}$, is required to be less than 4.5 \gevc\ 
	  for  the \EPPr\ and \EPPw\ final states.
\end{itemize}

In addition, particle ID vetoes are applied to specific selection channels.
For the $ehh$ decay modes, except for $eKK$, the 1-prong track must 
not be identified as an electron.
This requirement is useful to reduce possible four-fermion $eehh$ 
background events produced from two-photon or Bhabha-like processes.
For the decay modes with only pions
$(\EPPr, \EPPw, \MPPr, \MPPw)$ or the mixed kaon-pion modes
$(\EKPr, \EPKr, \EKPw, \MKPr, \MPKr, \MKPw)$,
the lepton candidate also must not pass the kaon criteria.

\subsection{\boldmath{\dEdM{}} OBSERVABLES}

To reduce backgrounds further,
candidate signal events are required to have
an invariant mass and total energy in the 3-prong
hemisphere consistent with a parent tau lepton.
These quantities are calculated from the observed track momenta 
assuming the corresponding lepton and hadron masses in each decay 
mode.
The energy difference is defined as 
$\Delta E \equiv E^{\mathrm{CM}}_{\mathrm{rec}} - E^{\mathrm{CM}}_{\mathrm{beam}}$,
where $E^{\mathrm{CM}}_{\mathrm{rec}}$ is the total energy of the tracks
observed in the 3-prong hemisphere and $E^{\mathrm{CM}}_{\mathrm{beam}}$
is the beam energy, both in the CM frame.
The mass difference is defined as
$\Delta M \equiv M_{\mathrm{rec}} - m_{\tau}$, where $M_{\mathrm{rec}}$ 
is the reconstructed invariant mass of the three tracks
and $m_{\tau}=1.777\gevcc$ is the tau mass \cite{bes}.

The signal distributions in the \dEdM\ plane are 
broadened by detector resolution and radiative effects.
The radiation of photons from the incoming \epem\ particles
before annihilation affects all decay modes, leading to a tail
at low values of \deltaE.
Radiation from the final-state lepton produces a tail
towards low values of \deltaM\ which
is more likely for electrons than muons.
Rectangular signal regions are defined separately for each 
decay mode as follows.
For all fourteen decay modes, the energy difference \deltaE\
must be in the range $[-100, +50] \mev$.
For the modes with muons, the mass difference \deltaM\ is required
to be in the range $[-20, +20] \mevcc$, while for the electron
modes the range is $[-30, +20] \mevcc$.

\begin{figure}
\resizebox{0.98\textwidth}{!}{%
  \includegraphics{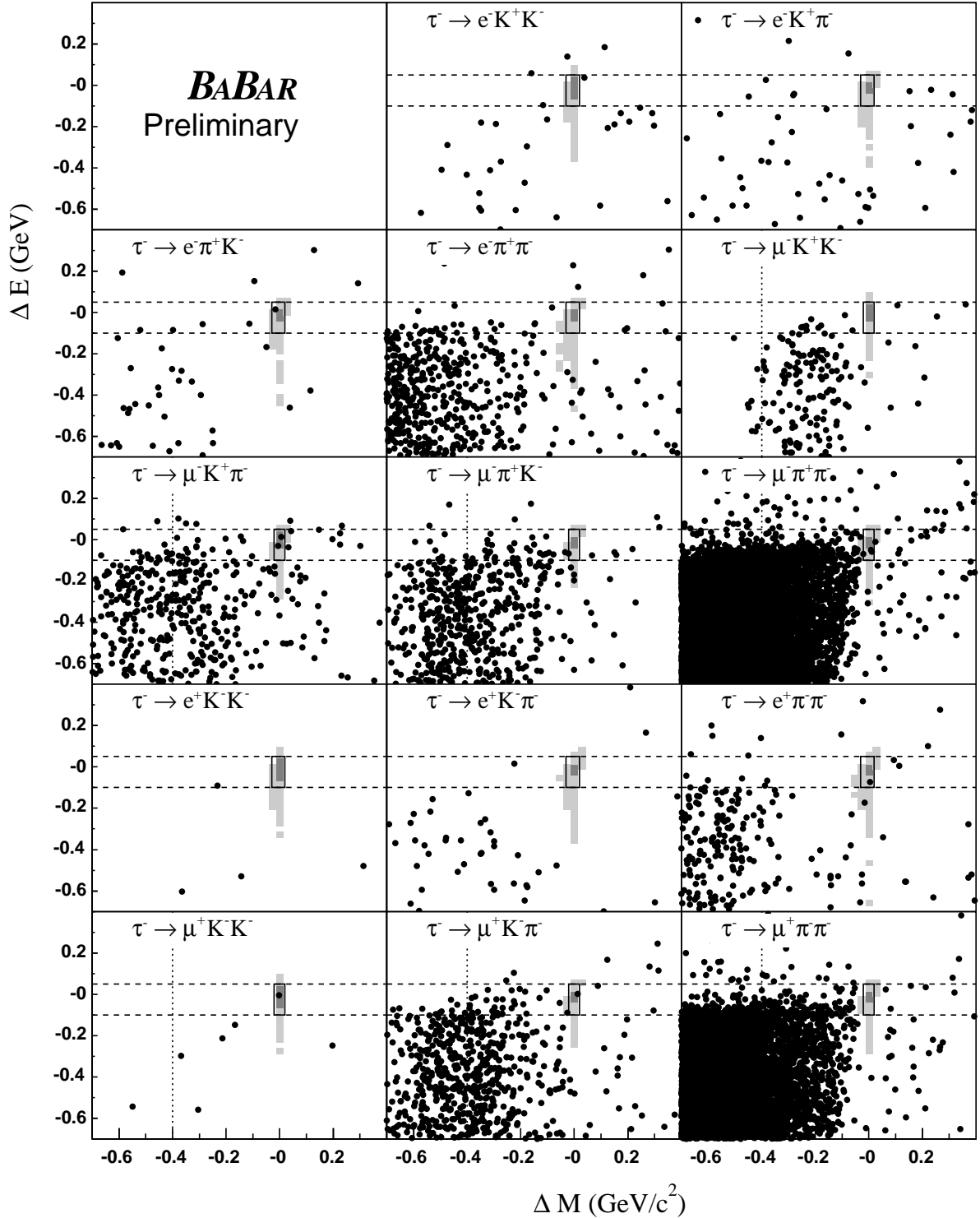} 
}
\caption{Observed data shown as dots in the \dEdM\ plane and 
the boundaries of the signal region for each decay mode.
The dark and light shading indicates contours containing
50\% and 90\% of the selected MC signal events, respectively.
The regions shown in Fig.~\ref{fig2} are indicated by dashed lines.
The vertical dotted lines show the lower $\Delta M$ boundary 
of the grand sideband (GS) region for the $\mu hh$ final states.
The other GS boundaries are the plot borders.}
\label{fig1}
\end{figure}

These signal region boundaries are chosen
to provide the smallest expected upper 
limits on the branching fractions in the background-only 
hypothesis.  These expected upper limits for the signal
box tuning are estimated using 
only Monte Carlo simulations, not candidate signal events.
Figure~\ref{fig1} shows the observed data for all
fourteen selection channels in the \dEdM{} plane,
along with the signal region boundaries 
and the expected signal distributions.
To avoid bias, a blind analysis procedure was adopted
with the number of data events in the signal region
remaining unknown until the selection criteria 
were finalized and all systematic studies had been performed.

\subsection{SELECTION EFFICIENCY}

The efficiency of the selection for signal events is
estimated with a MC simulation of LFV tau decays.
About 40\% of the MC signal events pass the initial 1-3 topology 
requirement.
From 20\% to 70\% of these preselected events pass the particle
ID criteria, depending upon the final state, and 70\% fall within
the signal region in the \dEdM\ plane.
The final selection requirements accept from 17\% to 27\% of these 
remaining signal MC events. 
The final efficiency for signal events to be found
in the signal region is shown in Table~\ref{tab:results} for
each decay mode and ranges from 2.1\% to 3.8\%.
This efficiency includes the 85\% branching fraction for 1-prong 
tau decays.

The particle ID efficiencies and misidentification 
probabilities are not taken from Monte Carlo, but rather measured 
using tracks in kinematically-selected data samples.
These values are parameterized as a function of particle momentum, 
polar angle, and azimuthal angle in the laboratory frame.
The control samples used to characterize the lepton ID performance
include radiative Bhabha, radiative \mumu, two-photon $\epem\ellell$, 
and $\jpsi\to\ellell$ events.
The control sample used to characterize the hadron ID performance
is the decay $\Dstar\to D^0\pi$, $D^0\to K\pi$.
These data-derived efficiencies are then used to determine the
probability that a simulated MC particle will be identified 
(or misidentified) as an electron, muon, pion, or kaon.

The lepton identification has been designed to give very low
mis-identification rates at the expense of some efficiency.
The electron ID is expected to be 81\% efficient in signal 
\lhh\ events with a mis-ID rate of 0.1\% for pions and 
0.2\% for kaons expected in generic \tautau\ events.
The muon ID is 44\% efficient for signal events, with
a mis-ID rate of 1.0\% for pions and 0.4\% for kaons.
The hadronic identification is designed to classify the
hadronic candidates as pions or kaons, but is not intended 
to distinguish hadrons from leptons.
Since the dominant backgrounds contain many pions (and some kaons),
stricter requirements on identifying hadrons do not 
improve the analysis.
The pion ID is 92\% efficient with a mis-ID rate of 12\% for
kaons, while the kaon ID is 81\% efficient with a 1.4\% mis-ID
rate for pions.
The pion ID criteria also has a very high mis-ID rate 
for muons (98\%) and electrons (38\%) which means that the
\MPPr\ channel is also rather sensitive to other LFV tau
decay modes like $\tau\to\mmm$.

\begin{table}
\begin{center}
\caption{Efficiency estimates, the number of expected background events (\Nbgd) in the signal region,
the number of observed events (\Nobs) in the signal region, and the 90\% CL 
upper limit for each decay mode.  
}
\begin{tabular}{lcccc}
\hline\hline
Mode & Eff. [\%] & \Nbgd  & $N_{obs}$ & UL at 90\% CL\\
\hline
\EKKr &$ 3.77 \pm 0.16 $&$ 0.22 \pm  0.06 $& 0 &$1.4 \cdot 10^{-7}$\\
\EKPr &$ 3.08 \pm 0.13 $&$ 0.32 \pm  0.09 $& 0 &$1.7 \cdot 10^{-7}$\\
\EPKr &$ 3.10 \pm 0.13 $&$ 0.14 \pm  0.06 $& 1 &$3.2 \cdot 10^{-7}$\\
\EPPr &$ 3.30 \pm 0.15 $&$ 0.81 \pm  0.15 $& 0 &$1.2 \cdot 10^{-7}$\\
\MKKr &$ 2.16 \pm 0.12 $&$ 0.24 \pm  0.08 $& 0 &$2.5 \cdot 10^{-7}$\\
\MKPr &$ 2.97 \pm 0.16 $&$ 1.67 \pm  0.32 $& 2 &$3.2 \cdot 10^{-7}$\\
\MPKr &$ 2.87 \pm 0.16 $&$ 1.04 \pm  0.20 $& 1 &$2.6 \cdot 10^{-7}$\\
\MPPr &$ 3.40 \pm 0.19 $&$ 2.99 \pm  0.42 $& 3 &$2.9 \cdot 10^{-7}$\\
\EKKw &$ 3.85 \pm 0.16 $&$ 0.04 \pm  0.04 $& 0 &$1.5 \cdot 10^{-7}$\\
\EKPw &$ 3.19 \pm 0.14 $&$ 0.16 \pm  0.06 $& 0 &$1.8 \cdot 10^{-7}$\\
\EPPw &$ 3.40 \pm 0.15 $&$ 0.41 \pm  0.10 $& 1 &$2.7 \cdot 10^{-7}$\\
\MKKw &$ 2.06 \pm 0.11 $&$ 0.07 \pm  0.10 $& 1 &$4.8 \cdot 10^{-7}$\\
\MKPw &$ 2.85 \pm 0.16 $&$ 1.54 \pm  0.28 $& 1 &$2.2 \cdot 10^{-7}$\\
\MPPw &$ 3.30 \pm 0.18 $&$ 1.46 \pm  0.23 $& 0 &$0.7 \cdot 10^{-7}$\\
\hline
\hline
\end{tabular}
\label{tab:results}
\end{center}
\end{table}

\subsection{BACKGROUND ESTIMATES}

There are two main classes of background remaining after
the selection criteria are applied: low multiplicity \qqbar{} events
(mainly continuum light-quark production) and SM \tautau{} events.
These background classes have distinctive distributions
in the \dEdM\ plane:
\qqbar{} events tend to populate the plane uniformly,
while \tautau{} backgrounds are restricted to negative 
values of both $\Delta E$ and $\Delta M$.
Backgrounds from Bhabha and \mumu events (which were important
for our \taulll\ analysis \cite{babar03}) are found to be negligible 
in the \taulhh\ decay modes.
The two-photon backgrounds are also found to be negligible, although
studies show they would look very similar to \qqbar\ backgrounds.

For each background class, a probability density function (PDF)
describing the shape of the background distribution in the \dEdM\
plane is determined by fitting an analytic function to the Monte
Carlo prediction as described in more detail below.
These two PDFs are then combined with coefficients determined by a
fit to the observed data in the \dEdM\ plane in a grand sideband
(GS) region which excludes the signal region.
The resulting function describes the event rate seen in the
GS region and is then used to predict the expected 
background rate in the signal region.

The GS region, shown in Fig.~\ref{fig1},
is defined as the rectangle bounded by 
$(-700\mevcc < \deltaM < 400\mevcc)$ for \ehh\ final states,
$(-400\mevcc < \deltaM < 400\mevcc)$ for \mhh\ final states,
and $(-700\mev < \deltaE < 400\mev)$ for both, excluding the
signal region.
For the \qqbar\ backgrounds, an analytic PDF is
constructed from the product of two PDFs $P_{M'}$ and 
$P_{E'}$, where
$P_{M'}(\Delta M')$ is a Gaussian and 
$P_{E'}(\Delta E') = (1-x/\sqrt{1+x^2})(1+a x+b x^2+c x^3)$ with 
$x=(\Delta E'-d)/e$\cite{opal00}.
The $(\Delta M', \Delta E')$ axes have been rotated slightly
from \dEdM\ to better fit the expected distributions.
The resulting \qqbar\ PDF is described by eight fit parameters,
including the rotation angle, which are determined by fits to MC
\qqbar\ background samples for each decay mode.
The \tautau\ PDF function $P_{M'}(\Delta M')$ is the sum of a Gaussian and
a bifurcated Gaussian, while the functional form of $P_{E'}(\Delta E')$
is the same as the \qqbar\ PDF above. 
To properly model the kinematic limit in tau decays, a similar
coordinate transformation is performed to get $\Delta M'$ and
$\Delta E'$, except that the axes are not required to remain
orthogonal.\footnote{The transformation is $\Delta M' = 
\cos\beta_1 \Delta M + \sin\beta_1 \Delta E$ and 
$\Delta E' = \cos\beta_2 \Delta E - \sin\beta_2 \Delta M$.
The coefficients $\beta_1$ and $\beta_2$ are also fit from the 
predicted MC background distributions.
}
This method reproduces well the slightly wedge-shaped background
distribution seen from tau decays in Figure~\ref{fig1} for the
$\mu\pi\pi$ final states.
In total there are 12 free parameters describing this PDF, and
all are determined by fits to MC \tautau\ samples, including the 
two coordinate transformation angles.

\begin{figure}
 \resizebox{\columnwidth}{!}{%
\includegraphics{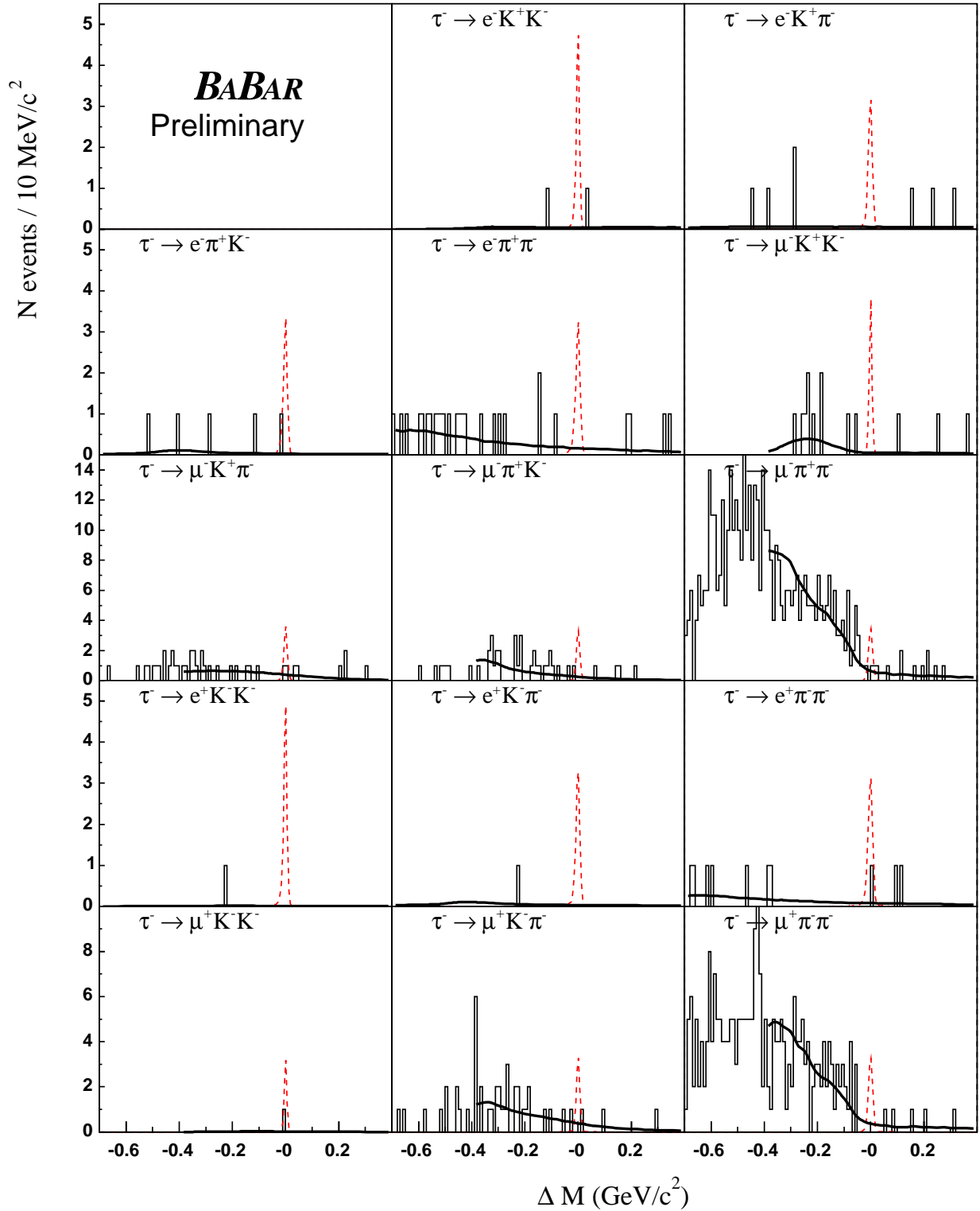}}
\caption{Distribution of $\Delta M$ for data (solid histogram) and
background PDFs (solid curves) for events with $\Delta E$ in the
range $[-100, +50] \mev$ as indicated in Fig.~\ref{fig1}.
Expected signal distributions are shown (dashed histogram)
for a branching fraction of \mbox{$5 \cdot 10^{-7}$}.}
\label{fig2}
\end{figure}

With the shapes of the two background PDFs determined, 
an unbinned maximum likelihood fit to the data in the GS region
is used to find the expected rate of each background type
in the signal region, as shown in Table~\ref{tab:results}.
The PDF shape determinations and background fits are performed
separately for each of the fourteen decay modes.
Figure~\ref{fig2} shows the data and the background PDFs
for values of $\Delta E$ in the signal range.

\section{SYSTEMATIC STUDIES}
\label{sec:Systematics}

The largest systematic uncertainty for
the signal efficiency is due to the uncertainty in
measuring the particle ID efficiencies.
This uncertainty\footnote{All uncertainties quoted in the 
text are relative.} is determined from the statistical precision
of the particle ID control samples, 
and ranges from $0.7\%$ for \EPPr\ to $3.8\%$ for \MKKr.
The modeling of the tracking efficiency contributes an additional 2.5\%
uncertainty, while the statistical uncertainty of the MC signal samples
ranges from 1\% to 2\%.
All other sources of uncertainty are found to be 
small, including the  
modeling of radiative effects, 
track momentum resolution, trigger performance, observables used in 
the selection criteria, and knowledge of the tau 1-prong branching 
fractions.
The efficiency has been estimated using a 3-body phase space model
and no uncertainty is assigned for possible model dependence.
The selection efficiency is found to be uniform within 
20\% across the Dalitz plane, provided
the invariant mass for any pair of particles from the LFV decay is 
less than $1.4 \gevcc$.

Since the background levels are extracted directly from the data,
systematic uncertainties on the background estimation are directly
related to the background parameterization and the fit technique used.
The finite data available in the GS region used to determine 
the background rates is the largest uncertainty and varies
from 14\% to 140\% depending upon the decay mode.
Additional uncertainties of 10\% are estimated by varying the fit 
procedure and changing the functional form of the background PDFs.
The uncertainty on the branching fraction of SM tau decays with
one or two kaons in the final state contributes 0-14\% to the
uncertainty on the estimated background.  Cross checks of the
background estimation were performed by considering the number of
events expected and observed in sideband regions immediately
neighboring the signal region for each decay mode as shown in
Table~\ref{tab:neighbor}.
Good agreement is seen between the observed data and background 
expectations.

\begin{table}[htbp]
\begin{center}
\caption{
The number of events expected (left) and observed (right) in the
regions just neighboring the signal region.
These "neighbor boxes" have the same dimension as the original
signal region for each decay mode, but have been offset by the
width or height of the signal region in the \dEdM\ plane.
}
\begin{tabular}{l|cc|cc|cc|cc}
\hline\hline
Mode       &\multicolumn{2}{|c|}{left} &\multicolumn{2}{|c|}{top}
&\multicolumn{2}{|c|}{right} &\multicolumn{2}{|c}{bottom} \\
\hline
  \EKKr & 0.22 &  0 & 0.05 &  1 & 0.24 &  1 & 0.30 &  0 \\
  \EKPr & 0.31 &  0 & 0.10 &  0 & 0.31 &  0 & 0.35 &  0 \\
  \EPKr & 0.15 &  0 & 0.05 &  0 & 0.13 &  0 & 0.18 &  0 \\
  \EPPr & 0.89 &  0 & 0.36 &  1 & 0.76 &  0 & 0.93 &  0 \\
  \MKKr & 0.30 &  1 & 0.04 &  0 & 0.21 &  0 & 0.33 &  0 \\
  \MKPr & 1.81 &  0 & 0.16 &  0 & 1.37 &  2 & 1.79 &  1 \\
  \MPKr & 1.30 &  1 & 0.14 &  0 & 0.84 &  0 & 1.09 &  3 \\
  \MPPr & 4.66 & 10 & 0.97 &  0 & 2.09 &  1 & 2.68 &  0  \\
  \EKKw & 0.05 &  0 & 0.00 &  0 & 0.03 &  0 & 0.05 &  0 \\
  \EKPw & 0.17 &  0 & 0.07 &  0 & 0.14 &  0 & 0.19 &  0 \\
  \EPPw & 0.43 &  0 & 0.13 &  0 & 0.37 &  0 & 0.43 &  1 \\
  \MKKw & 0.08 &  0 & 0.01 &  0 & 0.06 &  0 & 0.08 &  0 \\
  \MKPw & 1.87 &  2 & 0.23 &  0 & 1.26 &  0 & 1.83 &  0 \\
  \MPPw & 2.42 &  3 & 0.52 &  0 & 1.15 &  1 & 1.54 &  0 \\
\hline\hline
\end{tabular}
\label{tab:neighbor}
\end{center}
\end{table}

\section{RESULTS}
\label{sec:Physics}

The numbers of events observed (\Nobs) and the background expectations
(\Nbgd) are shown in Table~\ref{tab:results}, with no significant
excess found in any decay mode.  Upper limits on the branching
fractions are calculated according to $\BRul = \Nul/(2 \varepsilon \L
\sigma_{\tau\tau})$, where $\Nul$ is the 90\% CL upper limit for the
number of signal events when \Nobs\ events are observed with \Nbgd\
background events expected.  The values $\varepsilon$, $\L$, and
$\sigma_{\tau\tau}$ are the selection efficiency, luminosity, and
\tautau{} cross section, respectively.  The estimates of $\L = \lumi$
and $\sigma_{\tau\tau} = 0.89$ nb are correlated,\footnote{The
luminosity is measured using the observed \mumu rate, and the \mumu
and \tautau{} cross sections are both estimated with {\tt KK2f}.}  and
the uncertainty on the product $\L \sigma_{\tau\tau}$ is 2.3\%.  The
upper limits on the branching fraction have been calculated including
all uncertainties using the technique of Cousins and Highland
\cite{cousins92} following the implementation of Barlow
\cite{barlow02}.  The 90\% CL upper limits on the \taulhh\ branching
fractions, shown in Table~\ref{tab:results}, are in the range
\taulhhlimits.  These limits represent an order of magnitude
improvement over the previous experimental bounds \cite{cleolhh}.

\section{ACKNOWLEDGMENTS}
\label{sec:Acknowledgments}

We are grateful for the 
extraordinary contributions of our \pep2\ colleagues in
achieving the excellent luminosity and machine conditions
that have made this work possible.
The success of this project also relies critically on the 
expertise and dedication of the computing organizations that 
support \babar.
The collaborating institutions wish to thank 
SLAC for its support and the kind hospitality extended to them. 
This work is supported by the
US Department of Energy
and National Science Foundation, the
Natural Sciences and Engineering Research Council (Canada),
Institute of High Energy Physics (China), the
Commissariat \`a l'Energie Atomique and
Institut National de Physique Nucl\'eaire et de Physique des Particules
(France), the
Bundesministerium f\"ur Bildung und Forschung and
Deutsche Forschungsgemeinschaft
(Germany), the
Istituto Nazionale di Fisica Nucleare (Italy),
the Foundation for Fundamental Research on Matter (The Netherlands),
the Research Council of Norway, the
Ministry of Science and Technology of the Russian Federation, and the
Particle Physics and Astronomy Research Council (United Kingdom). 
Individuals have received support from 
CONACyT (Mexico),
the A. P. Sloan Foundation, 
the Research Corporation,
and the Alexander von Humboldt Foundation.

\end{document}